# Impact of Polymeric precursor and Auto-combustion on the Structural, Microstructural, Magnetic, and Magnetocaloric Properties of $La_{0.8}Sr_{0.2}MnO_3$

Afaak Lakouader[1,2,*], Youness Hadouch[1,2], Daoud Mezzane[1,2], Valentin Laguta[3,4] , M'barek Amjoud[1], Voicu O. Dolocan[5], Nikola Novak[6], Lahoucine Hajji[1], El Hassan Choukri[1], Anna Razumnaya[6], Abdelhadi Alimoussa[1], Zdravko Kutnjak[6] ,Igor A. Luk'yanchuk[2,7], and Mimoun El Marssi[2]

[1] IMED-Lab, Cadi Ayyad University, Marrakesh, 40000, Morocco

[2] LPMC, University of Picardy Jules Verne, Amiens, 80039, France

[3] Institute of Physics AS CR, Cukrovarnicka 10, Prague, 16200, Czech Republic

[4] Institute for Problems of Materials Science, National Ac. of Science, Krjijanovskogo 3, Kyiv 03142, Ukraine

[5] Aix-Marseille University, CNRS, IM2NP UMR7334, Marseille Cedex 20, F-13397, France

[6] Condensed Matter Physics Department, Jožef Stefan Institute, Jamova Cesta 39, 1000, Ljubljana, Slovenia

[7] Department of Building Materials, Kyiv National University of Construction and Architecture, Kyiv, 03680, Ukraine

* Corresponding author: Laboratory of Innovative Materials, Energy and Sustainable Development (IMED), Cadi- Ayyad University, Faculty of Sciences and Technology, BP 549 Marrakech, Morocco.
E-mail address: afaak.lakouader@gmail.com (A.Lakouader)

## Abstract:

In this work, $La_{0.8}Sr_{0.2}MnO_3$ (LSMO) nanopowders are synthesized using two different methods: Pechini (LSMO-PC) and auto-combustion (LSMO-AC). Nanoparticle sizes, structural, magnetic, and magnetocaloric properties were determined and compared. The X-ray diffraction confirms the coexistence of two phases; rhombohedral symmetry with space group *R-3c* and orthorhombic symmetry with space group *Pbnm*, with the rhombohedral phase dominating. The scanning electron microscope images show that LSMO-PC has larger nanoparticle sizes (~ 495 nm) than LSMO-AC (~195 nm). The samples exhibit ferromagnetic properties with distinct hysteresis loops and Curie temperatures 340 K and 290 K for LSMO-PC and LSMO-AC respectively. The variation of the magnetic entropy was measured indirectly using the Maxwell approach with increasing magnetic field. For LSMO-PC it reaches a maximum $-\Delta S_M$=1.69 J/kg.K at 340 K and $\Delta H$= 5 T. The associated adiabatic temperature change $\Delta T_M$ is 1.04 K. While LSMO-PC demonstrates superior magnetic and magnetocaloric properties, LSMO-AC displays significant magnetocaloric thermal stability.

The obtained values make LSMO-PC and LSMO-AC promising candidates for eco-friendly room-temperature magnetocaloric applications.

**Keywords:** Manganite, Ferromagnetic, Magnetocaloric effect, Pechini, Auto-combustion, Mean-field model, 3D Heisenberg model.

## 1 Introduction

Mixed valence manganite exhibits intriguing electrical and magnetic behaviors such as a transition from a metal to an insulator [1], colossal magnetoresistance (CMR) [2], charge order phase transitions [3], and magnetocaloric effect (MCE) [4]. The MCE is a phenomenon represented by an isothermal magnetic change entropy or an adiabatic temperature change when a magnetic field is applied to or withdrawn from magnetic material. This method has been extensively suggested as an alternative cooling method to traditional gas compression or Peltier-based technique.

Manganite perovskites with the chemical formula $A_{1-x}B_xMnO_3$, where A is a trivalent rare earth ion ($La^{3+}$, $Pr^{3+}$, $Sm^{3+}$) and B is a divalent alkaline earth ion ($Ca^{2+}$, $Sr^{2+}$, $Ba^{2+}$) have garnered a lot of interest due to their remarkable electrical and magnetic properties [5]. The magnetic and electrical properties of manganite are controlled by the interaction between electron-phonon coupling resulting in the Jahn-Teller effect and the double exchange (DE) mechanism [6], [7]. Electrical transport, magnetic properties, and transition from ferromagnetic (FM) to paramagnetic (PM) are affected by many factors such as doping level *x*, $Mn^{4+}/Mn^{3+}$ ions ratio, interaction between $Mn^{4+}/Mn^{3+}$, oxygen ions internal stress and structural defects [8].

The remarkable magnetic and electrical properties of manganite perovskites have attracted the attention of both experimental and theoretical researchers. Typical manganite perovskites involve $La_{1-x}Sr_xMnO_3$ (LSMO) [9], $La_{1-x}Ca_xMnO_3$ (LCMO) [10], and $La_{1-x}Ba_xMnO_3$ (LBMO) [11]. Pure $LaMnO_3$ is an insulating antiferromagnetic material (AFM) with a Neel temperature at $T_N$ = 140 K and the electronic configuration $Mn^{3+}$ [12]. However, when $La^{3+}$ ions are replaced by divalent cations (D= $Ca^{2+}$, $Sr^{2+}$, $Ba^{2+}$) in $La_{1-x}D_xMnO_3$, a mixed valence electronic configuration of $Mn^{3+}$ and $Mn^{4+}$ is produced. The super-exchange interaction results from the formation of $Mn^{3+}$-O-$Mn^{3+}$ and $Mn^{4+}$-O-$Mn^{4+}$ bonds, whereas a DE mechanism is possible with $Mn^{3+}$-O-$Mn^{4+}$ bonds. Above a certain critical level of *x* doping concentration, the DE mechanism dominates over the super exchange interaction, and $La_{1-x}D_xMnO_3$ undergoes a PM to FM transition with an important diminution in resistivity. One of the main mechanisms controlling the magnetic and transport properties of manganite is the DE mechanism. The variation of the Mn-O bond length and Mn-O-Mn bond angle can facilitate orbital and spin ordering mechanisms through spin-lattice and spin-orbit coupling, which contribute to the electrical and magnetic properties [9], [13]. Sr-substituted manganites are of interest because $La_{1-x}Sr_xMnO_3$ (LSMO) displays high electrical conductivity, magnetic properties, and high Curie temperature. LSMO is a spin-canted insulator for $0 < x < 0.1$ and an FM insulator for $0.1 < x < 0.12$. However, when the composition is in the range of 0.12 - 0.6, the LSMO is FM metallic coupled to the CMR effect [14], [15], and for x > 0.6, a purely AFM and insulating state develops within a nearly cubic structure [14], [16].

In study of the MCE in perovskite manganite compounds, most researchers concentrated on the dopant-operating effects, dopant concentration, and sintering temperature. Lu et al. showed for $La_{0.67}Sr_{0.33}MnO_3$ that with the increase of the calcination temperature, the magnetic entropy $\Delta S_M$ increases. The obtained value of $-\Delta S_M$ at magnetic field change ($\Delta H$) of 1.5 T were 0.32, 1.3, and 1.74 J/kg.K for sintering temperature of 950 °C, 1150 °C, and 1400 °C, respectively [17]. Szewczyk et al. reported the MCE in $La_{1-x}Sr_xMnO_3$ manganite with $x$ = 0.120, 0.135, 0.155, 0.185, and 0.200 [18] and demonstrated that except for the x = 0.120 composition, the MCE increases with increasing Sr-doped content. For composition x = 0.20, the magnetocaloric temperature change $\Delta T$=4.15 K at $\Delta H$ = 7 T was reported [18]. Similarly, Mira et al. observed an elevated value of magnetic entropy $-\Delta S_M$ =1.5 J/kg.K at 370 K for $\Delta H$ = 1 T in $La_{0.65}Sr_{0.35}MnO_3$ [19]. Phan et al. reported a large MCE in $La_{0.7}Ca_{0.3-x}Sr_xMnO_3$ single crystals with $0.05 \leq x \leq 0.25$ at room temperature. The $-\Delta S_M$ reaches a value of 10.5 J/kg.K at 275 K for $\Delta H$ = 5 T and $x$ = 0.05 [20].

The MCE of manganites synthesized using various methods is rarely reported and compared. The electrical and magnetic properties of manganite perovskites are strongly influenced by particle size and morphology, which vary depending on the synthesis method. For preparing of manganite perovskites wet chemical routes are most often used due to better homogeneity, chemical purity, and stoichiometric compositions that can be achieved. These wet chemical methods include sol-gel process [21], pyrophoric reaction process [22], hydrothermal synthesis [23], pyrolysis [24], co-precipitation method [25], and sol-gel-based polymeric precursor route [26]. Because these techniques do not require costly and toxic reagents are processed at low reaction temperatures with short reaction durations, they are economical and can be used in large-scale applications. Auto-combustion and Pechini routes are two of the most convenient and effective wet chemical methods for synthesizing magnetic oxides due to their simplicity and good-control of the morphology of particles. For powder synthesis with the auto-combustion technique a low reaction temperature 80-100 °C and a fast reaction time are characteristic [27]. The Pechini technique provides polycondensation between metallic ions and hydroxycarboxylic ethylene glycol and citric acid to produce a polymeric matrix that limits particle growth [33]. Furthermore, the immobilization of metal chelates in the rigid polymer network using this method prevent metal ion segregation during polymer decomposition at high temperatures [28].

In this study, $La_{0.8}Sr_{0.2}MnO_3$ (LSMO) powders were synthesized using Pechini and auto-combustion routes. The effect of both synthesis methods on the structural, microstructural, magnetic, and magnetocaloric properties of LSMO was examined and discussed. In addition, an indirect method based on Maxwell's thermodynamic relation is used to study the magnetocaloric properties.

## 2 Materials and Methods

The $L_{0.8}Sr_{0.2}MnO_3$ powders were synthesized by two different methods: Pechini and auto-combustion. All starting materials were analytical grade and used without further purification.

### 2.1 Preparation of LSMO powder by Pechini method

Stoichiometric amounts of the nitrate precursor reagents ($La(NO_3)_3{\cdot}6H_2O$, $Mn(NO_3)_2{\cdot}4H_2O$ and $Sr(NO_3)_2$) were dissolved in water. Then, citric acid was added as a chelating agent to the solution at a molar ratio of 4:1 with respect to LSMO. After stirring the resulting mixture for 30 min, an appropriate amount of ethylene glycol (EG) as a dispersant, was added under stirring. The solution was then heated under constant stirring at 80 °C to obtain a viscous gel. To remove the excess ethylene glycol produced during the polymerization and oligomerization stages, the final solution was heated for 24 h at 100 °C. The dried gel was then calcined at 700 °C for 4 h to stabilize the crystalline structure of the desired powder abbreviated as LSMO-PC.

### 2.2 Preparation of LSMO powder by auto-combustion

Stoichiometric amounts of the nitrate precursor reagents ($La(NO_3)_3{\cdot}6H_2O$, $Mn(NO_3)_2{\cdot}4H_2O$, and $Sr(NO_3)_2$ ) were dissolved in distilled water under continuous stirring, and then, ethylene glycol and citric acid were added as chelating agent and fuel, respectively. The resulting solution was stirred and heated for 1 h at 80 °C resulting in the formation of the gel. Next, the gel was heated for 1 h at 200 °C until completely transformed into fine powder. Finally, the obtained powder was annealed at 700 °C for 4 h to remove any remaining organic materials and to form the desired compound designated as LSMO-AC.

### 2.3 Characterization

The powder phase structure was determined by X-ray diffraction (XRD), analysis using the Cu-Kα radiation with $\lambda$~1.540598 Å at Panalytical X-Pert Pro spectrometers. The measurements were performed at room temperature in a range of 2θ from 20° to 80° with a step size of 0.02°. Rietveld refinement analysis was realized by using FullProf software. The surface characteristics of LSMO powders were qualitatively scanned by the scanning electron microscope (SEM) coupled with energy-dispersive X-ray spectroscopy (EDS).

The magnetization as temperature function was obtained by using a physical property measurement system (PPMS), DynaCool Quantum Design apparatus operating in the temperature range of 2–400 K and magnetic fields up to 1 T. The isotherm magnetization curves were measured by using the vibrating sample magnetometer (VSM) method at different temperatures 250 K- 400 K with steps of 10 K, under a magnetic field variation of $\pm$50 kOe. Ferromagnetic resonance (FMR) was measured using the Bruker EMX plus and Bruker E580 EPR spectrometers at X microwave band (9.4 GHz) at 290 to 380 K. The magnetocaloric temperature change was determined by the indirect method from recorded (M-H) hysteresis as a function of temperature.

## 3 Results and discussion

### 3.1 Structural and microstructural properties

**Fig.1 (a)** and **(b)** show the room temperature XRD patterns of $L_{0.8}Sr_{0.2}MnO_3$ powders synthesized by Pechini (LSMO-PC) and auto-combustion (LSMO-AC) methods, respectively. Both samples have complete single-phase perovskite structure, without crystalline impurity phase. The sharp diffraction peaks indicate good crystallinity [29]. Zhang et al. reported that the orthorhombic symmetry is often dominant for $La_{1-X}Sr_XMnO_3$ with $x < 0.2$. However, for $0.25 < x < 0.5$, LSMO exhibits a transition to rhombohedral symmetry [30]. Diffractograms with Rietveld refinement show that both LSMO samples crystallized in mixed crystalline phases. The estimated crystalline phases that best fit the experimental results are orthorhombic symmetry with space group *Pbnm* and rhombohedral symmetry with space group *R-3c*. The most prevalent crystallographic phase appears to be the rhombohedral phase, with ~ 92.76 % and ~ 98.10% for LSMO-PC and LSMO-AC, respectively. Structural parameters and phase composition are gathered in **Table 1**. The lattice constants of LSMO-PC and LMSO-AC are practically identical for both the rhombohedral and orthorhombic structures. The structures of the orthorhombic and rhombohedral phases are significantly different since they are characterized by different tilt systems of the $MnO_6$ octahedra [31]. The $Mn^{3+}$ and $Mn^{4+}$ ions form a deformed zigzag pattern in the rhombohedral structure (see **Fig.1 (c)**). Still, in the orthorhombic structure, they are located in a straight line (see **Fig.1 (d)**), resulting in a slightly different angle for the $Mn^{3+}$-O-$Mn^{4+}$ bonds (see **Table 1**). When the angle approaches 180°, the orbitals of the $Mn^{3+}$ and $Mn^{4+}$ ions are aligned better, and their overlap is improved, which results in a stronger DE mechanism [32]. Conversely, the DE mechanism becomes weaker when the angle deviates from 180° because the overlap is reduced. Due to the larger portion of orthorhombic phase in LSMO-PC compared to LSMO-AC, we expect that higher percentage of $Mn^{3+}$-O-$Mn^{4+}$ bonds to be near 180° resulting in a stronger DE mechanism. In addition, the unit cell volume of the rhombohedral structure in both LSMO-PC and LSMO-AC compounds is ~ 33% higher than that of the orthorhombic structure. A smaller cell volume of both crystal phases, rhombohedral and orthorhombic, is observed for LSMO-PC compared to LSMO-AC (see **Table 1**). At this stage, we may expect that LSMO-PC should exhibit better magnetic properties since the cell volume is related to the bond length of Mn-O in the crystal lattice. The unit cell volume decreases with decreasing the Mn-O bond distance, which improve orbital overlap, hence and consequently the DE mechanism efficiency [33]. Accordingly, with ~ 7.24% of orthorhombic phase, LSMO-PC could engage in the DE mechanism more successfully, exhibiting good magnetic and magnetocaloric properties [34].

Using Scherrer's formula, the average crystallite size is calculated from the XRD peaks as [35]:

$$d = {}^{K\lambda}\!/_{\beta\cos\theta} \cdot \qquad (1)$$

Here $d$ is the size of the particle, $K$ is known as the Scherer's constant (K=0.94), $\lambda$ is X-ray wavelength, β is the corrected full-width half maxima of the XRD peaks, and θ is the diffraction angle. The average crystallite size is 31.54 nm for LSMO-PC and 23.62 nm for LSMO-AC.

The SEM images are displayed in the inset of **Fig.2**. It is interesting to note that the shape of the particles depends on the synthesis method. LSMO-PC depicts agglomerated particles of different sizes, shapes, with a wide distribution of size, and porosity between the agglomerates. LSMO-AC, on the other hand, has a granular morphology and the particles tend to aggregate to form a sheet-like material. Some grains have a complex layer structure in form of parallel and equidistant bands. The LSMO-AC surface is homogeneous with fewer pores. The sample of LSMO-AC contains small grains and very close contact between grain boundaries compared to LSMO-PC. The average particle size values are around 495 nm and 195 nm for LSMO-PC and LSMO-AC, respectively.

The elemental composition is estimated using EDS analysis shown in **Fig.2**. Qualitative analysis reveals the presence of the metal elements La, Sr, and Mn. The presence of carbon is due to the metallization process. **Table 1** shows the atomic percentages of the elements present in the samples. The obtained values agree with the expected ratios. The oxygen content was found to be 2.94 and 2.905 atoms per cell for LSMO-PC and LSMO-AC, respectively. The variations may result from various synthesis techniques. When using the Pichini method, the reaction environment can be precisely controlled to assure better oxygen retention [28], and the finished product may have a higher oxygen content than materials made via autocombustion. In autocombustion, the exothermic process produces heat, and the quick gas release might result in the oxygen being ejected [27].

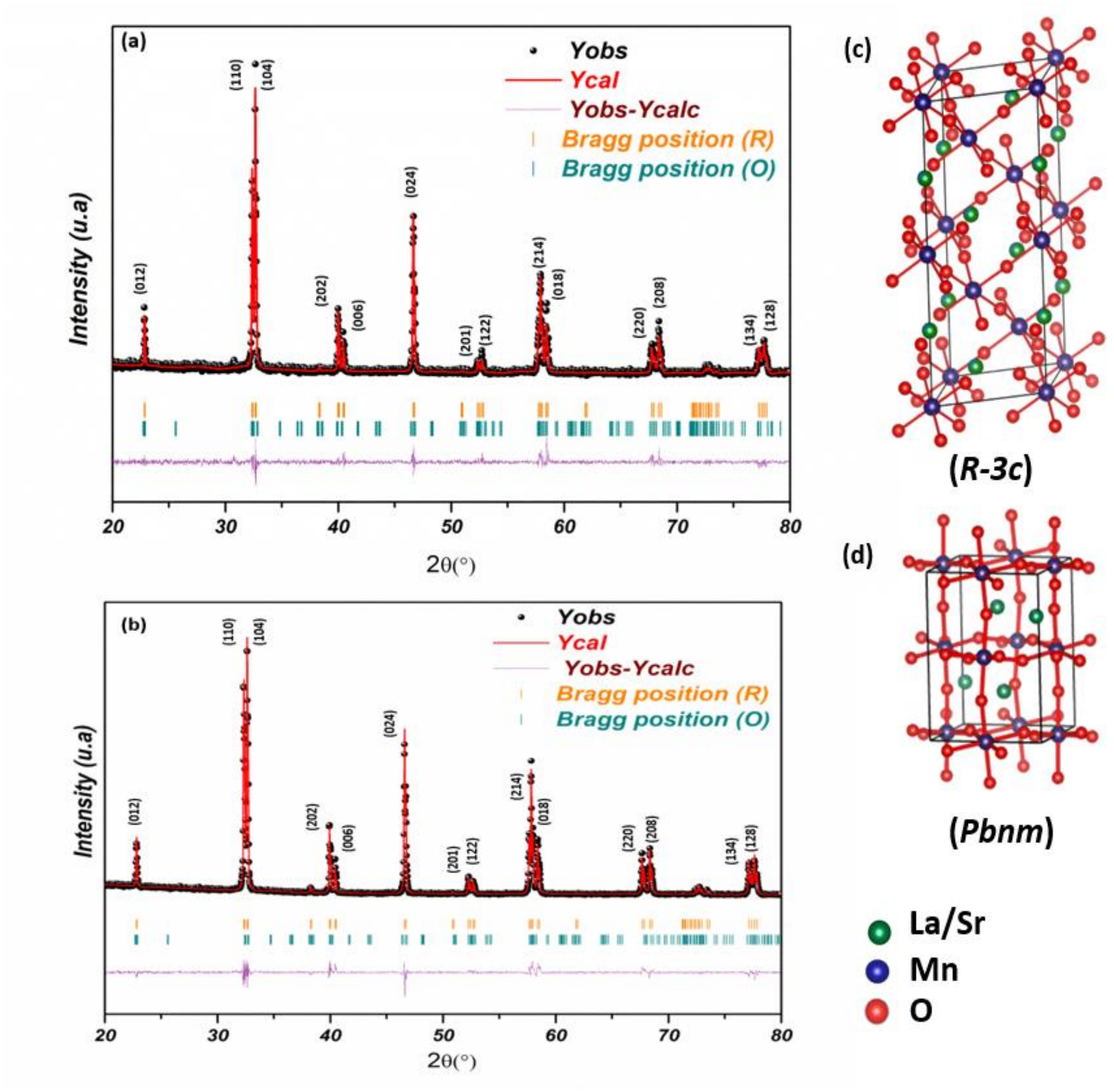

Figure 1: XRD patterns and Rietveld refinement of LSMO samples (a) LSMO-PC (b) LSMO-AC and (c) Rhombohedral structure (R-3c), (d) Orthorhombic structure (Pbnm).

Table 1: Structural and microstructural parameters and the elemental composition of LSMO-PC and LSMO-AC.

| | ***LSMO-AC*** | | ***LSMO-PC*** | |
|---|---|---|---|---|
| ***Structure*** | ***Rhombohedral*** | ***Orthorhombic*** | ***Rhombohedral*** | ***Orthorhombic*** |
| ***Group space*** | R -3 c | P b n m | R -3 c | P b n m |

<table>
<tr><td rowspan="5">Unit cell parameters</td><td>a(Å)</td><td colspan="2">5.52598</td><td colspan="2">5.53739</td><td colspan="2">5.53431</td><td colspan="2">5.52090</td></tr>
<tr><td>b(Å)</td><td colspan="2">5.52598</td><td colspan="2">5.45478</td><td colspan="2">5.53431</td><td colspan="2">5.47798</td></tr>
<tr><td>c(Å)</td><td colspan="2">13.37262</td><td colspan="2">7.81860</td><td colspan="2">13.37643</td><td colspan="2">7.83111</td></tr>
<tr><td>α=β (°)</td><td colspan="2">90</td><td colspan="2">90</td><td colspan="2">90</td><td colspan="2">90</td></tr>
<tr><td>γ (°)</td><td colspan="2">120</td><td colspan="2">90</td><td colspan="2">120</td><td colspan="2">90</td></tr>
<tr><td colspan="2">Volume cell ($Å^3$)</td><td colspan="2">354.81</td><td colspan="2">236.84</td><td colspan="2">353.64</td><td colspan="2">236.16</td></tr>
<tr><td colspan="2">Bond distance (Å) $\langle d_{Mn-O} \rangle$</td><td colspan="2">1.96</td><td colspan="2">1.95</td><td colspan="2">1.95</td><td colspan="2">1.93</td></tr>
<tr><td colspan="2">Bond angle (°) Mn-O1-Mn</td><td colspan="2">153.15</td><td colspan="2">166.35</td><td colspan="2">159.65</td><td colspan="2">169.93</td></tr>
<tr><td colspan="2">Bond angle (°) Mn-O2-Mn</td><td colspan="2">-</td><td colspan="2">165.02</td><td colspan="2">-</td><td colspan="2">165.02</td></tr>
<tr><td colspan="2">Phase composition (wt %)</td><td colspan="2">98.10</td><td colspan="2">1.90</td><td colspan="2">92.76</td><td colspan="2">7.24</td></tr>
<tr><td rowspan="3">Reliability factors (%)</td><td>χ2</td><td colspan="4">1.98</td><td colspan="4">1.08</td></tr>
<tr><td>Rp</td><td colspan="4">13.2</td><td colspan="4">5.79</td></tr>
<tr><td>Rwp</td><td colspan="4">16.2</td><td colspan="4">7.67</td></tr>
<tr><td colspan="2">Density by XRD ( $g/cm^3$)</td><td colspan="2">5.72</td><td colspan="2">6.91</td><td colspan="2">5.75</td><td colspan="2">7.05</td></tr>
<tr><td colspan="2">Average crystalline sizes (nm)</td><td colspan="4">23.62</td><td colspan="4">31.54</td></tr>
<tr><td colspan="2">Average particle sizes (nm)</td><td colspan="4">195</td><td colspan="4">495</td></tr>
<tr><td></td><td>Element</td><td>La</td><td>Sr</td><td>Mn</td><td>O</td><td>La</td><td>Sr</td><td>Mn</td><td>O</td></tr>
</table>

| | | | | | | | | | |
|---|---|---|---|---|---|---|---|---|---|
| ***Elemental composition by EDS*** | ***At (%)*** | 17.7 | 3.2 | 21 | 58.1 | 17.6 | 3.5 | 20.1 | 58.8 |

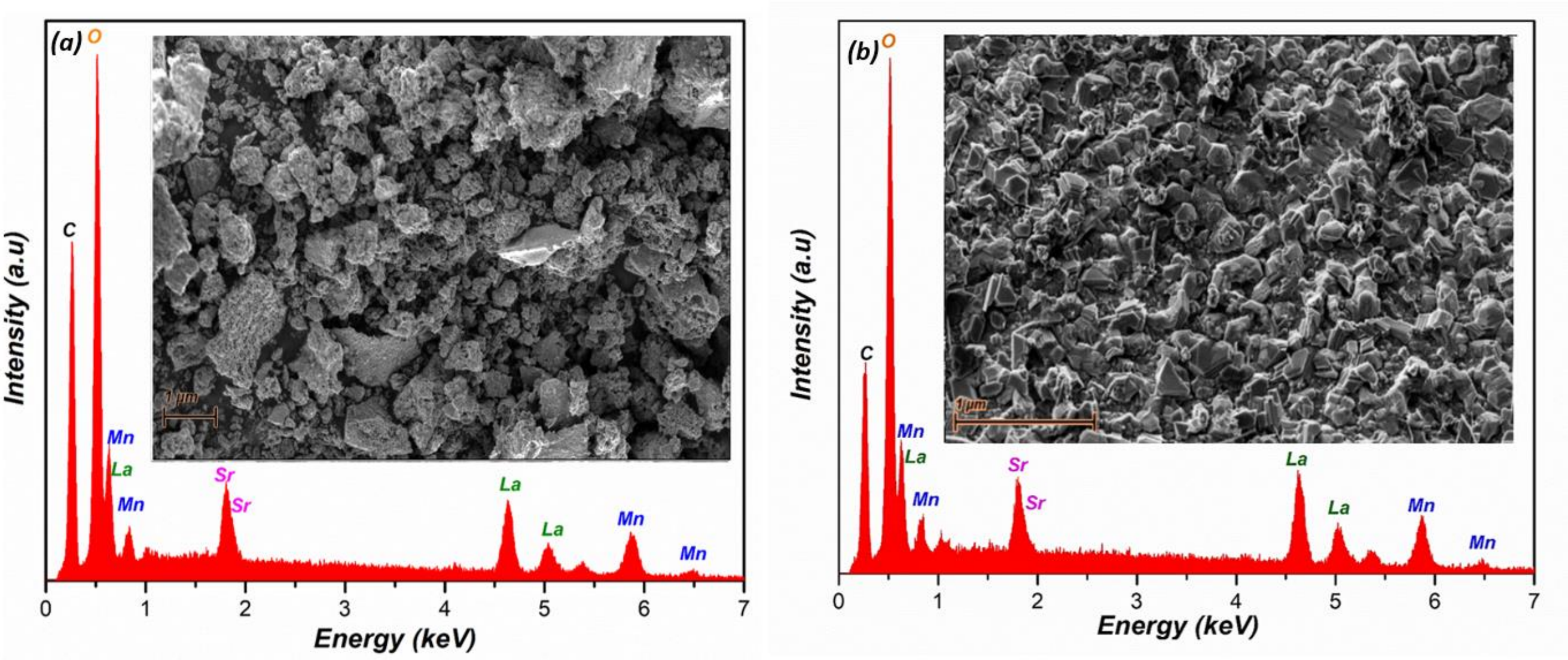

Figure 2: The EDS of LSMO samples (a) LSMO-PC (c) LSMO-AC. Inset: SEM images.

### 3.2 Magnetic properties

The temperature variation following field-cooled cooling (FCC) and field-cooled warming (FCW) protocols for the dc magnetization was measured for both LSMO-PC and LSMO-AC samples with a 1T in-plane magnetic field. There is no thermal hysteresis present between FCC and FCW sweeps (not shown), which ensures that the MCE is nearly temperature reversible [39]. Furthermore, the absence of thermal hysteresis in M(T) indicates that the ferromagnetic to paramagnetic phase transition for both samples is of the second order [49]. The FCC M(T) curves are displayed in **Fig.3 (a)** and **(c)** for LSMO-AC and LSMO-PC respectively. At 2 K, the saturation magnetization is of 60.82 emu/g for LSMO-PC and of 56.60 emu/g for LSMO-AC. As the temperature increases, the magnetization decreases, revealing a ferromagnetic to paramagnetic transition [35]. The Curie temperature $T_C$ determined from the minimum of the *dM/dT* curve (shown in the inset of panels (a) and (c)) is 340 K and 290 K for LSMO-PC and LSMO-AC, respectively. This difference is due to the DE interaction mechanism between $Mn^{3+}$-$Mn^{4+}$, which is affected by the crystalline phase, microstructure, unit cell volume, Mn-O-Mn bond angle, and Mn-O bond length [30], [36], [37]. LSMO-PC is more likely to take part in the double exchange process than LSMO-AC due to its higher proportion of orthorhombic phase (~ 7.24%), which leads to stronger ferromagnetic interactions and a higher Curie temperature. Moreover, the Mn-O-Mn bond angle is slightly smaller while the Mn-O bond length and unit cell volume are slightly larger in LSMO-PC compared to LSMO-AC sample (**Table 1**). As a result, the DE mechanism becomes weaker and the degree of Jahn-Teller distortion in LMSO-AC increases, which lowers $T_C$ and reduce the ferromagnetic properties [10]. The difference in magnetic properties between the two samples is greatly influenced by the microstructure and can be explained by the average grain size. The magnetization value increase as the average particle size increases [30]. Smaller averaged grain size results in the formation of more grain boundaries, which act as layers of magnetic disorder [38], and cause the formation of more broken bonds at the surface [39] which leads to a lowering of the magnetization value. On the other hand, the interaction of the magnetic properties at the surface and core regions is responsible for the magnetic response observed in magnetization curves for different particle sizes, as reported in other publications [40], [41]. It should be noticed that as particle size grows, the saturation magnetization rises. This behavior was discussed by considering an ideal inner core surrounded by an outer shell, and factors such vacancies, oxygen non-stoichiometry, and stress were all considered and accounted for [42]. As the particle size decrease, the outer shell increase (with thickness *t*), which causes the saturation magnetization for smaller particles to decrease [37]. This outer shell is named as the dead layer, which does not have any spontaneous magnetization. The thickness of the dead layer can be estimated using the following formula [43]:

$$t \approx {}^{d}\!/_{2}\left(1 - \left({}^{M_s}\!/_{M_{s0}}\right)^{1/3}\right) \quad (3)$$

where *t* is the dead layer thickness, *d* is the grain size, $M_s$ is the saturation magnetization of the nanopowders, and $M_s$ is the saturation magnetization the conventional bulk. The *t* is 3.1 nm and 3.6 nm for LSMO-PC and LSMO-AC, respectively. A decrease in the magnetic exchange energy results from an increase in the inter-core spacing between two nearby particles as the particle size decreases and the shell thickness t increases. Several factors result in the decrease of magnetization including contamination on the surface of grain, breakdown of Mn-O-Mn bonds, the creation of different arrangement of atoms and molecules at the surface and the presence of dislocations at grain boundaries [44]. These magnetization reduction effects are specific to the outer layer [42] and were reported in previous studies [45].

As shown in the inset of **Fig.3 (a)**, the minima in the dM/dT vs temperature curve for the LMSO-AC sample, is very large spreading over 120 K, with a first minima at 208 K and second minima at 290 K. This suggests a weak ferromagnetic coupling and that short-range ferromagnetic correlations may exists above $T_C$. To corroborate the existence of short-range order above $T_C$, the inverse susceptibility is plotted in **Fig.3 (b)** and **(d)** as a function of temperature for both samples. As observed in panel (b), for LMSO-AC, a downturn in the inverse susceptibility appears around 350 K, which is also an indication of ferromagnetic correlation above $T_C$. The linear relation which represents the Curie-Weiss law ($\chi = \chi_0/(T-T_{CW})$) was fitted to the experimental values above 370 K and the obtained Curie-Weiss temperature $T_{CW}$ was 312 K, larger than $T_C$, pointing to long-range correlations. However, the downturn in $\chi^{-1}(T)$ is usually a signature of the short-range ferromagnetic order in the paramagnetic phase called Griffiths phase [46]. The Griffiths phase (GP) is usually related to quenched disorder in a system and was observed experimentally in manganites[47]–[49], double perovskites systems[50], [51], and intermetallic systems[52]. Above the Griffiths temperature $T_G$, the systems becomes paramagnetic, while in the region $T_C < T < T_G$ a non-analytic relation between the magnetization and magnetic field exists due to magnetic clusters with short-range order embedded in the paramagnetic background. In the GP, the inverse susceptibility follows a power law $\chi^{-1} \approx (T - T_C^R)^{1-\lambda}$, where $\lambda$ measures the deviation from Curie-Weiss law ($\lambda$ is positive and inferior to 1) and $T_C^R$ is the critical temperature of ferromagnetic clusters above which the susceptibility diverges. The $\lambda$ value was calculated from the $\chi^{-1}$ plot against the reduced temperature ($T/ T_C^R -1$) in a log-log scale as shown in **Fig.4** for both the GP region and paramagnetic region. In the GP region, $\lambda \approx 0.812$ using $T_C^R = T_{CW}$ and $T_G = 370$ K, which indicates an important GP phase develops in LSMO-AC sample. In the paramagnetic phase (above $T_G$), $\lambda \approx 0.003$ that points to a pure paramagnetic phase (which corresponds to $\lambda = 0$). The magnetism of samples in the PM region is due to the formation of FM clusters of $Mn^{3+}$–$Mn^{4+}$ DE pairs containing more than one Mn ion [53].

To further analyze the ferromagnetic-paramagnetic transition, the critical analysis around $T_C$ was carried out. Sufficiently close to $T_C$, where critical fluctuation are important, the ferromagnetic properties show power law behavior in ($T$ - $T_C$) with critical exponents that specify a universality class[54]. The critical exponents for the order parameter (magnetization M), the conjugate magnetic field H and susceptibility $\chi$ are: $M \approx \varepsilon^{\beta}$ ($\varepsilon \geq 0$), $M = H^{1/\delta}$ ($\varepsilon = 0$) and $\chi \approx |\varepsilon|^{\gamma}$ ($\varepsilon < 0$), where $\varepsilon = (1 - T/T_C)$ is the reduced temperature. The critical exponents are not all independent but related through the scaling relation $\delta = 1 + \gamma/\beta$ [55]. In the critical region, the equation of state can be expressed as: $(H/M)^{\gamma} = a(T-T_C) - bM^{1/\beta}$, which allows to calculate the critical exponents from the M(H) isotherms.

**Fig.5 (a)** displays the M(H) isotherms for LSMO-AC sample, for temperatures between 250 K and 400 K, with a step of 10 K. The panel (b) shows the transformed isotherms into Arrott plots ($M^2$ *vs* H/M), which correspond to the mean field values of the critical exponents ($\beta$ = 0.5, $\gamma$ = 1, $\delta$ = 3) [56]. The positive slope of the plots points to a second order phase transition (Banerjee criterion) [57]. The isotherms should form a set of straight lines in high field around $T_C$, with the isotherm at $T_C$ passing through origin. From the mean field model, the $T_C$ is estimated at 289 K very close to the value obtained from *dM/dT* minimum. In panels (c) – (e), we plotted the modified Arrott plots (MAP) for the 3D Heisenberg model ($\beta$ = 0.365, $\gamma$ = 1.386, $\delta$ = 4.8), 3D Ising model ($\beta$ = 0.325, $\gamma$ = 1.241, $\delta$ = 4.82) and the mean field tricritical model ($\beta$ = 0.25, $\gamma$ = 1, $\delta$ = 5). For the LSMO-AC sample, the mean field model works best, indicating that a type of long-range ferromagnetic correlations are present in our system. We determined the critical exponent $\delta$ from the fit of the M(H) isotherm at $T_C$ and obtained $\delta$ = 3.40. Using the MAP analysis, we determined that $\beta$ is close to 0.41 and $\gamma$ is close to 0.96, implying that the critical exponents lie between the mean-field model (long-range interaction) and the 3D Heisenberg model (short-range interaction) for LSMO-AC.

Contrary to LSMO-AC, the minima in the *dM/dT* vs temperature curve for the LMSO-PC sample display the usual steep valley at 340 K (inset **Fig.3 (c)**). The downturn in the inverse susceptibility is less obvious, with a downturn around 354 K which correspond to $T_C^R$ while $T_G$ is very close to 400 K (maximum temperature available in our measurements). Fitting the inverse susceptibility with the power law above 390 K gives a critical exponent $\lambda \approx 0.2$, which implies that LSMO-PC is not completely paramagnetic at this temperature and a disordered Griffiths-like phase still appears. The same MAP analysis was applied to determine the critical exponents and is detailed in **Fig.6**. From the mean-field model (panel (b)), the Curie temperature is determined to be 336 K, close to the one determined from *dM/dT*. From comparing between different models presented in panels (c) to (e), the mean field model applies best for LSMO-PC. The critical exponents calculated from MAP are $\delta$ = 2.72, $\beta$ is close to 0.6 and $\gamma$ is close to 1.03. The $\beta$ value is above the mean field one, but very similar to Ca-doped nanocrystalline manganite [58]. Therefore, both LSMO-AC and LSMO-PC samples are close to conventional universality class presenting a small deviation from the mean-field values in opposite direction. This could be related with the disorder present in the system due to grain boundaries and grain size [59], [60].

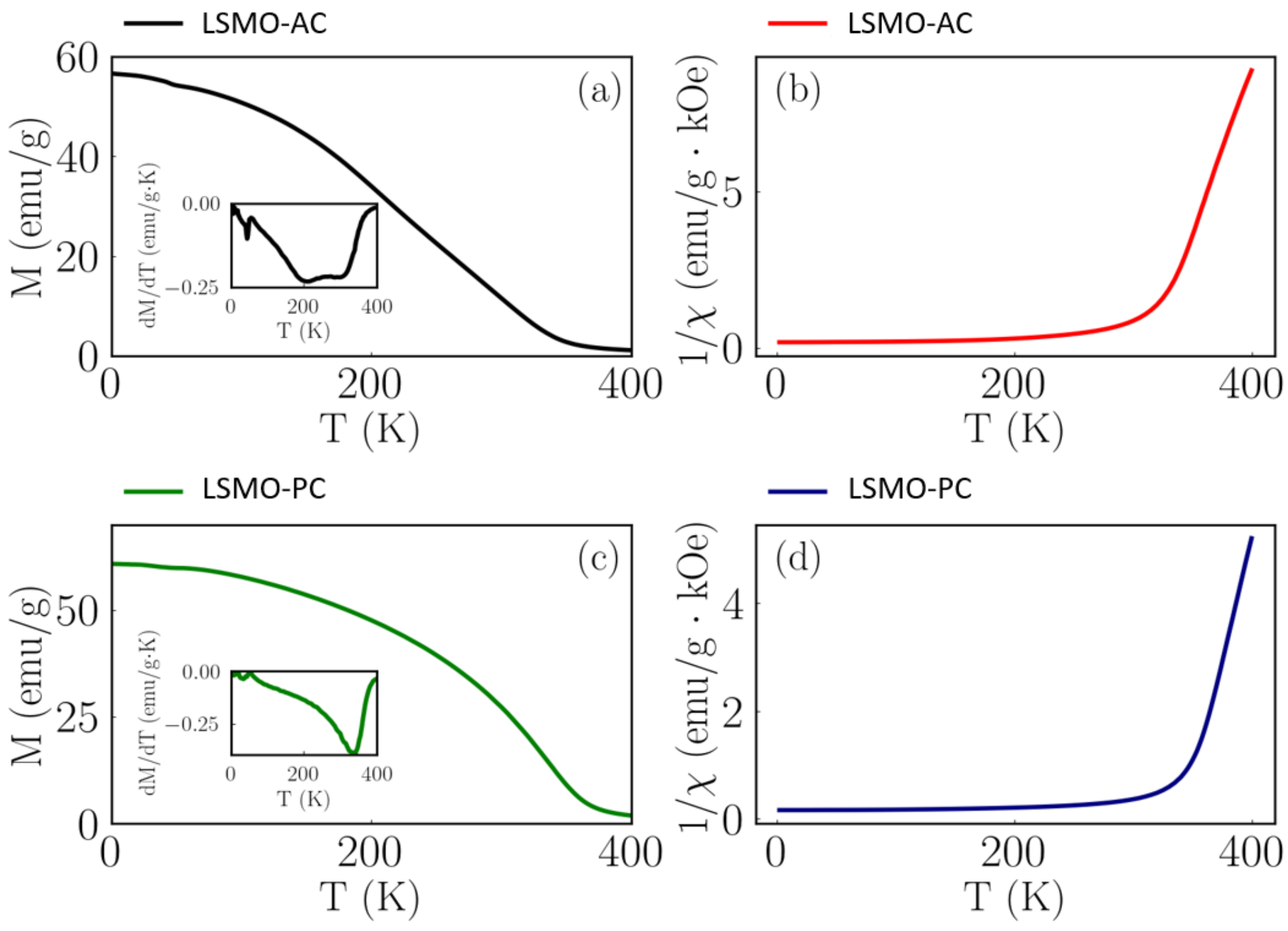

Figure 3: Temperature dependence of the magnetization for LSMO-AC sample (a) and and for LSMO-PC sample (c) measured in 1T in-plane field following field cooled cooling conditions. Inverse dc susceptibility of LSMO-AC (b) and of LSMO-PC (d) as a function of temperature. The inset in (a) and (c) shows the temperature variation of dM/dT.

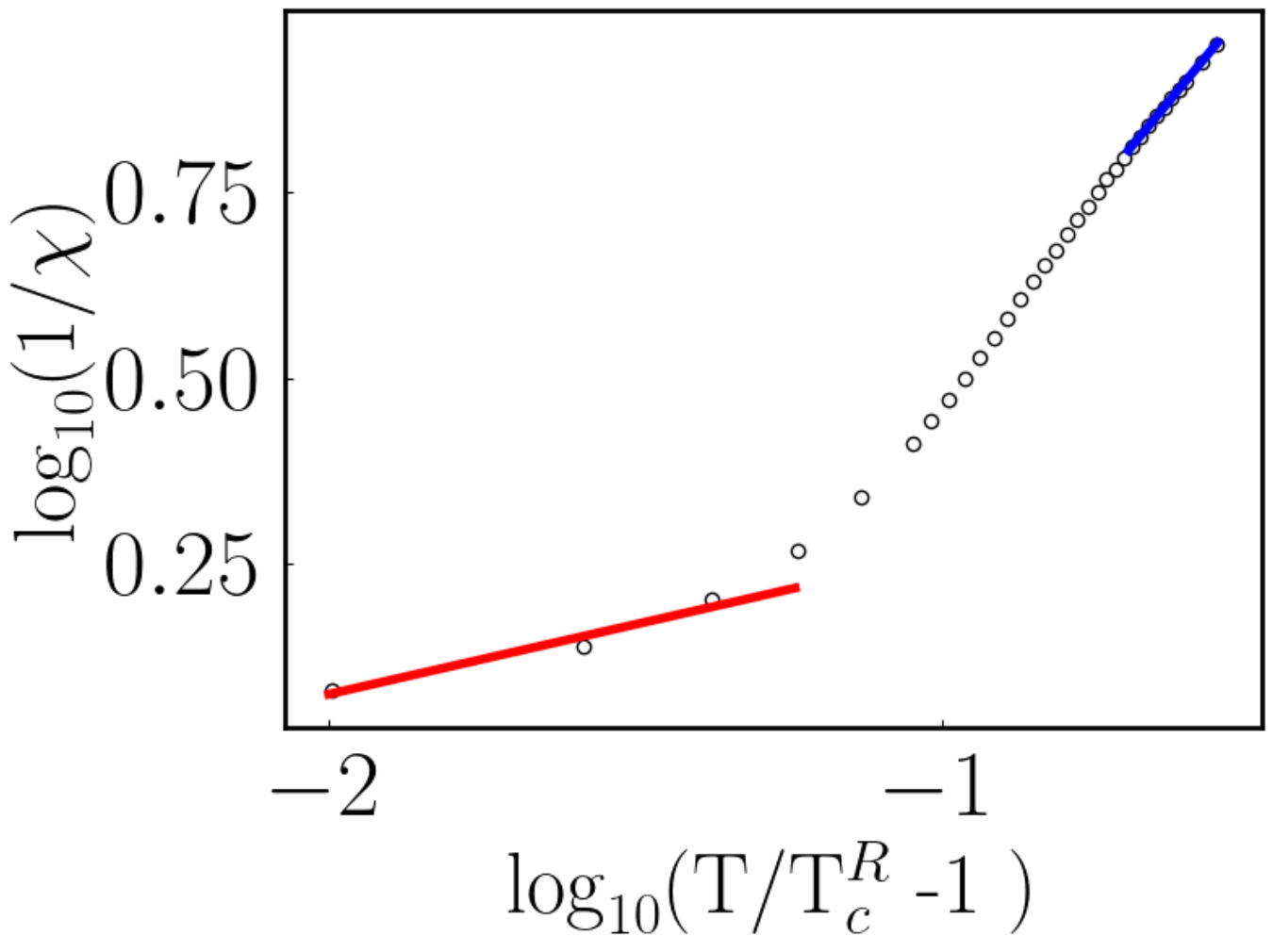

Figure 4: Log-log plot of the inverse susceptibility of LSMO-AC as a function of the reduced temperature. The lines indicate power law fits in the Griffith and paramagnetic regions.

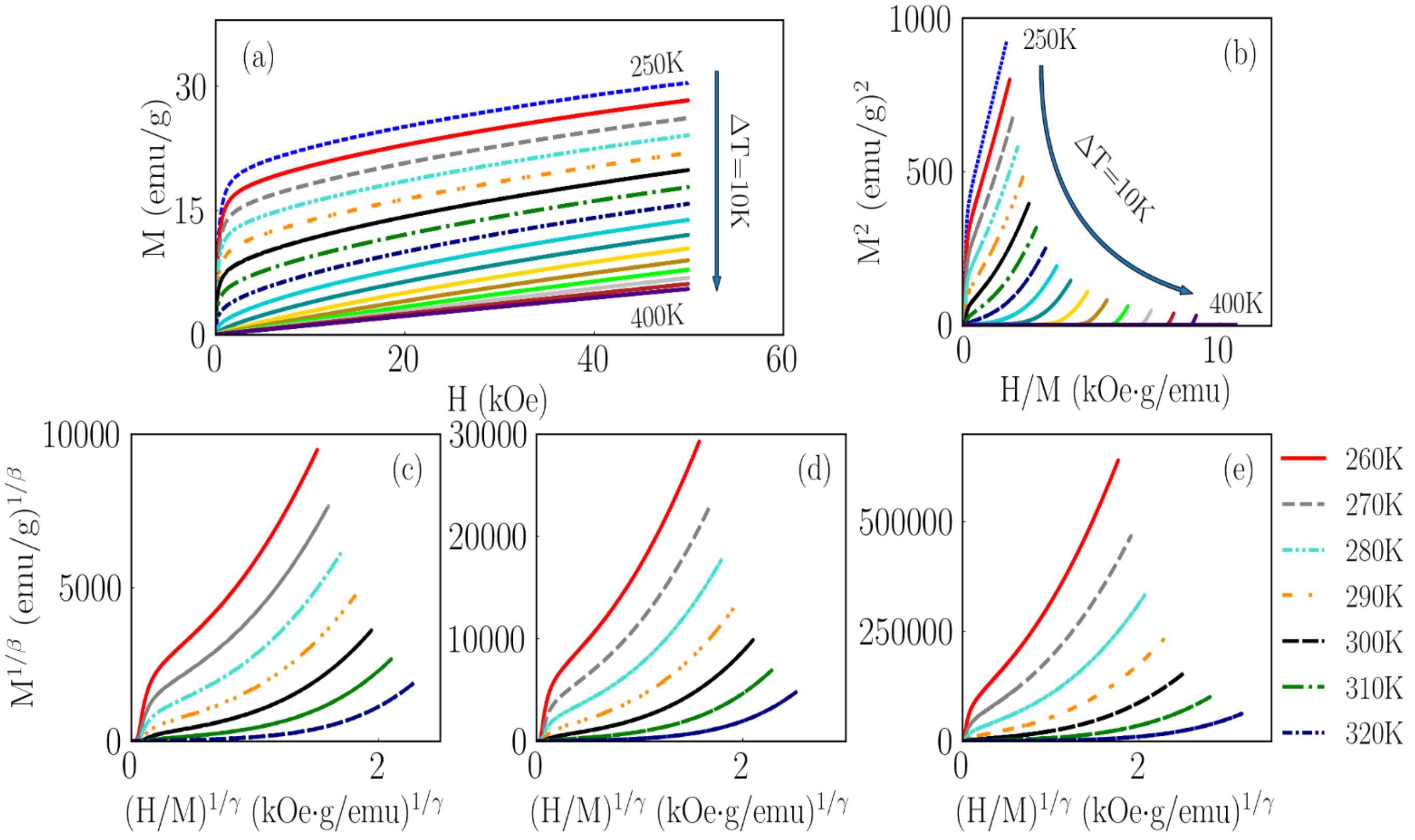

Figure 5: (a) M(H) isotherms of LSMO-AC measured at different temperatures from 250K to 400K. (b) Arrott plot ($M^2$ vs. H/M) of the isotherms shown in (a). The modified Arrott plots isotherms are shown in (c) – (e) using 3D Heisenberg, 3D Ising and tricritical mean-field model respectively.

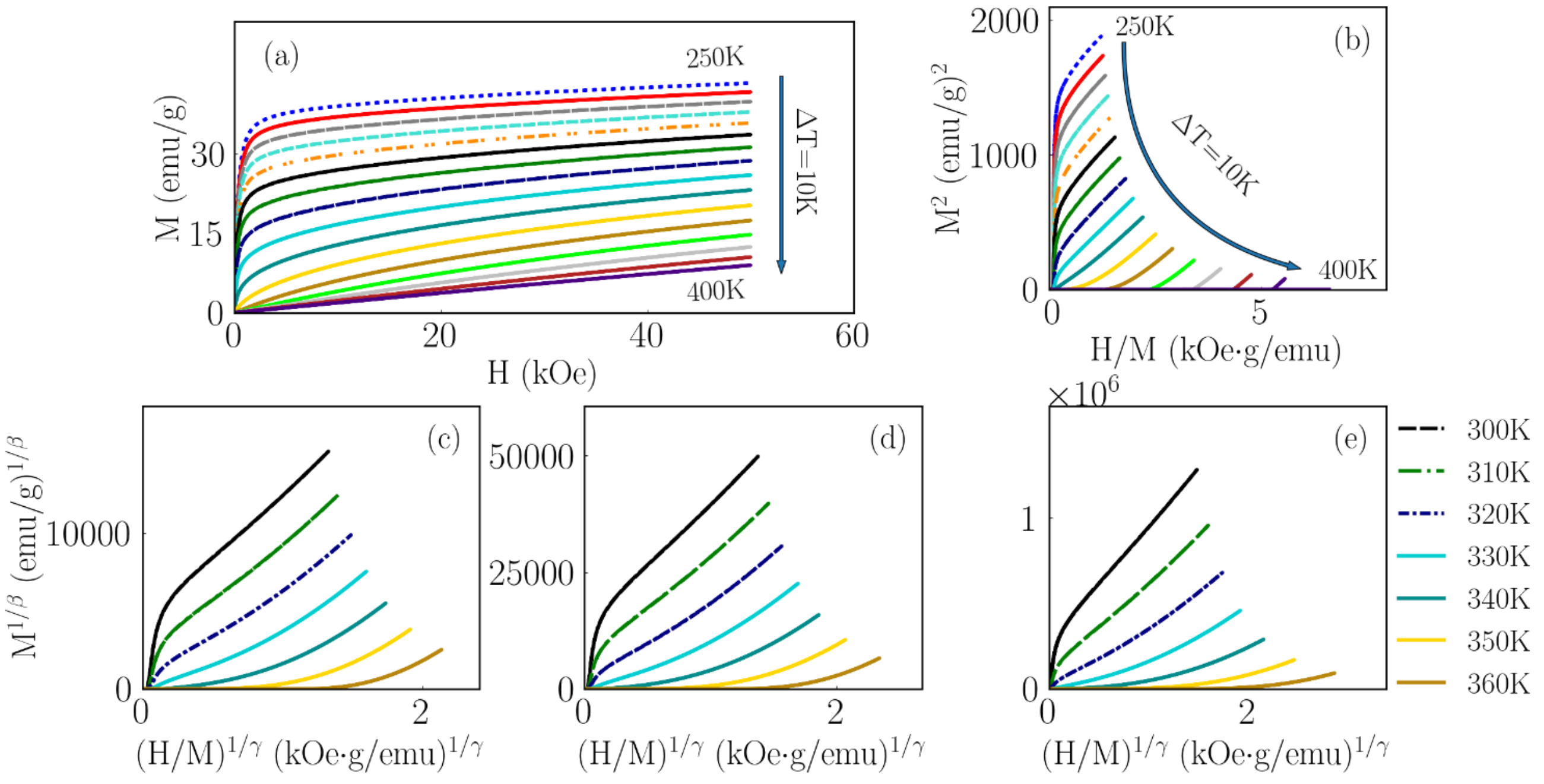

Figure 6: (a) M(H) isotherms of LSMO-PC measured at different temperatures from 250K to 400K. (b) Arrott plot ($M^2$ vs. H/M) of the isotherms shown in (a). The modified Arrott plots isotherms are shown in (c) – (e) using 3D Heisenberg, 3D Ising and tricritical mean-field model respectively.

### 3.3 Ferromagnetic resonance

Ferromagnetic resonance (FMR) is one of the most powerful experimental techniques in studying the magnetic properties [61]. It allows for the determination of magnetocrystalline anisotropy, magnetic moment, Curie and phase transition temperatures, and relaxation mechanisms of the magnetization. The FMR is explored to study the paramagnetic to ferromagnetic phase transition. **Fig.7** shows FMR spectra in the temperature interval from 295 to 370 K for both LSMO-AC and LSMO-PC powders. To confront with the broadness of the FMR powder spectra the original first derivatives of the signal measured by spectrometer were integrated to obtain absorption spectra that have simpler shape. In the paramagnetic phase, due to strong exchange interaction among $Mn^{4+}$ (S=3/2) and $Mn^{3+}$ (S=2) spins, the magnetic resonance spectrum consists of one spectral line of the Lorentz shape (see **Fig.7**, upper spectrum) whose intensity follows the Curie-Weiss law (**2**). The resonance field $H_r$ is determined by the average g-factor of $Mn^{4+}$ and $Mn^{3+}$ ions and is ≈2.0. As the temperature is decreased below the PM-FM phase transition temperature, the long-range magnetic ordering is established, resulting in the appearance of spontaneous magnetization and magnetocrystalline anisotropy field $H_A$. This leads to the shift of the resonance field and broadening of the spectral line as the anisotropy field is strongly anisotropic. **Fig.8** presents the temperature dependence of the resonance field $H_r$ and the linewidth, measured as the peak-to-peak distance of the first derivative spectrum. The linewidth in the PM phase decreases with decreasing temperature due to the spin-phonon relaxation mechanism at high temperatures. However, it starts to increase at phase transition temperature $T$c allowing precise determination of the PM-FM phase transition temperature. $T$c was 345 K and 330 K for LSMO-PC and LSMO-AC, respectively. Approximately the same $T$c values, albeit with lower precision are obtained from the shifts of resonance field, see panel b in **Fig. 8**. The obtained results corroborate the temperature-dependent magnetization measurement; namely, $T$c is higher in LSMO-PC than that in LSMO-AC. On the other hand, the $T$c = 290 K obtained from the temperature dependence of magnetization in LSMO-AC (**Fig.3**) is substantially lower than $T$c = 330 K obtained from FMR measurements. The difference in $T_C$ temperatures may be related with larger diffusion of the PM-FM phase transition in the powder prepared by the auto-combustion method as compared to the LSMO-PC powder. Formation of the Griffiths phase is not excluded as well.

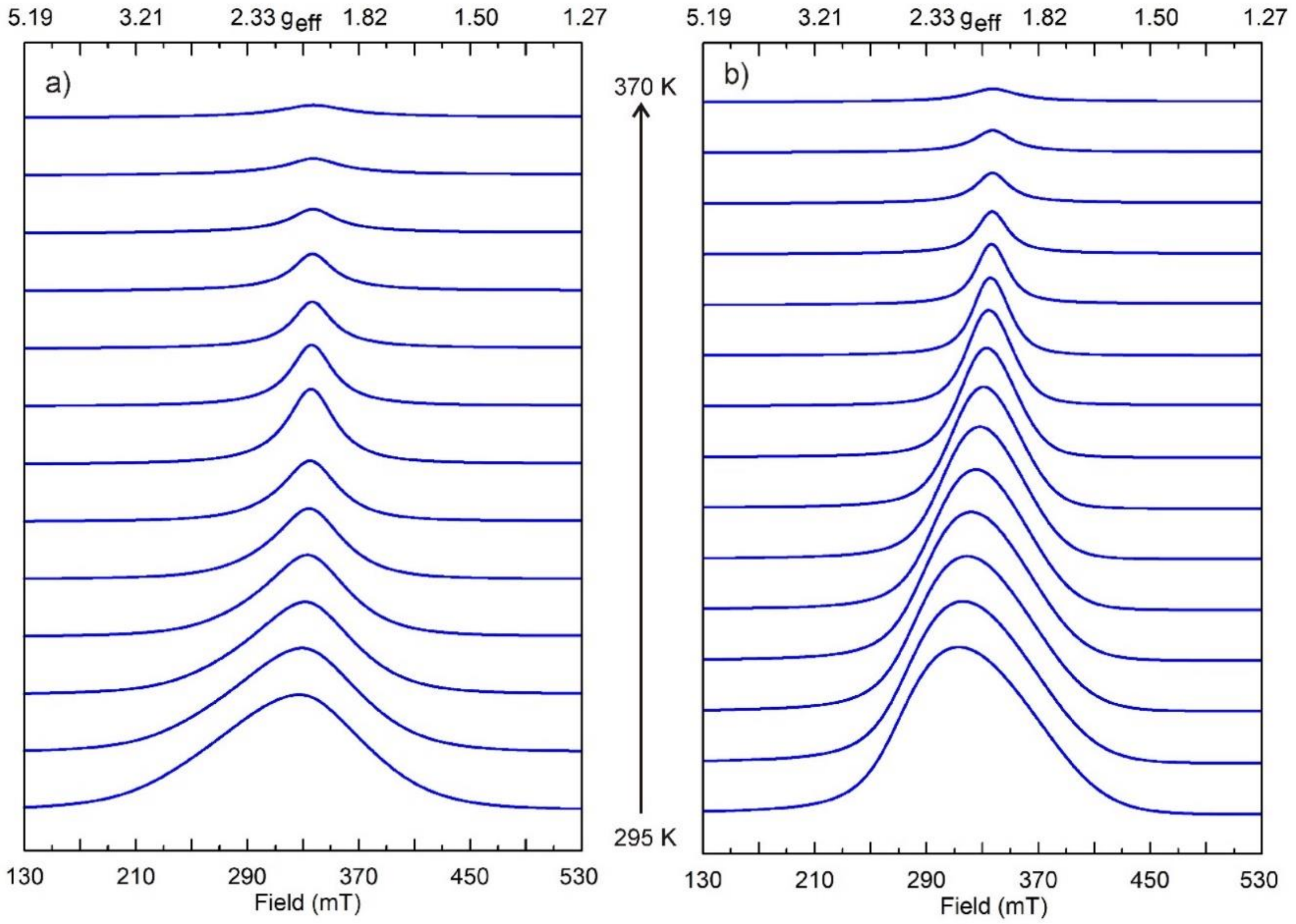

Figure 7: Temperature dependence of FMR absorption spectra measured for (a) LSMO-AC and (b) LSMO-PC.

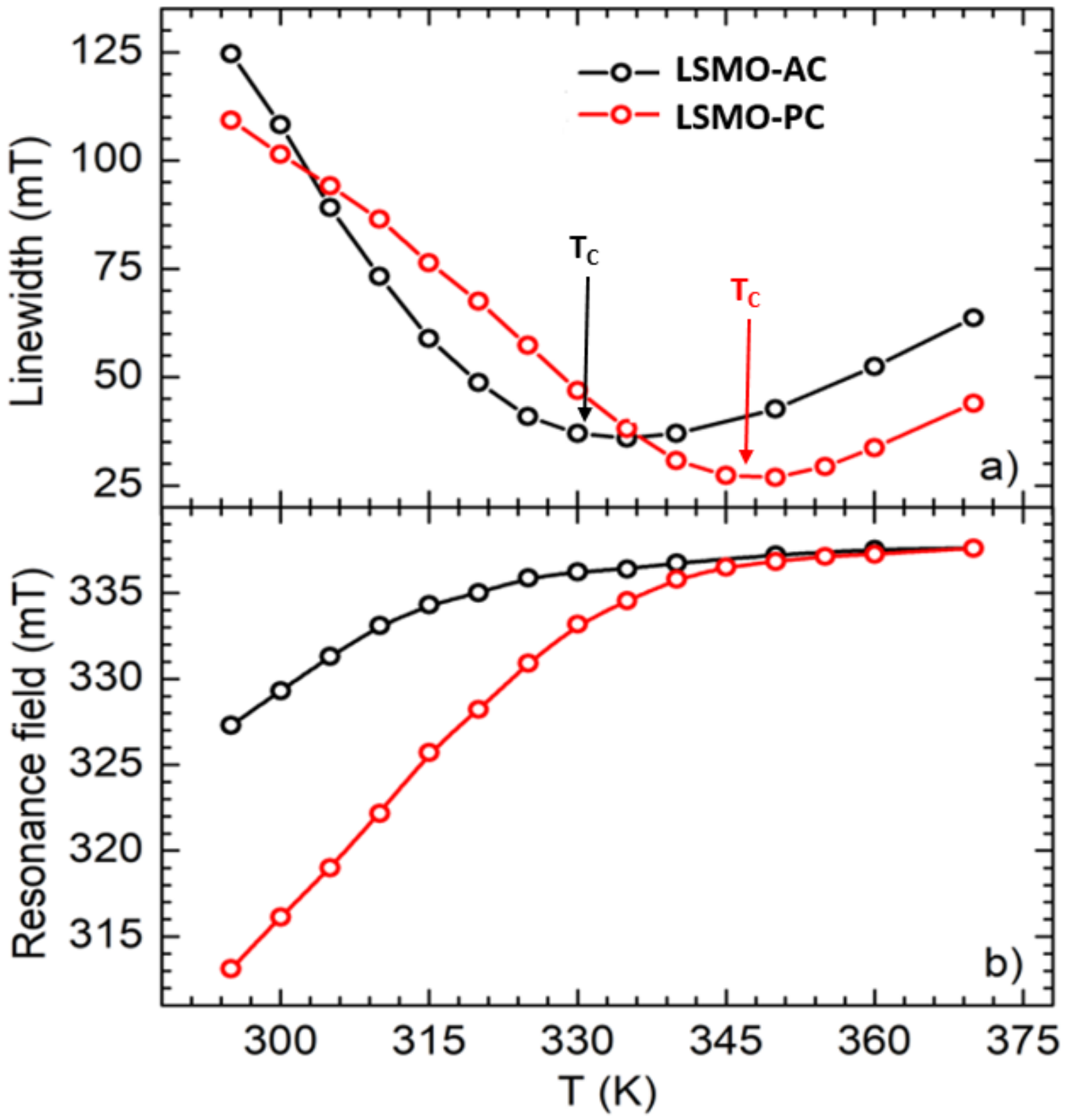

Figure 8: (a) Temperature dependence of linewidth (peak-to-peak distance of first derivative spectrum) and (b) FMR field measured for LSMO-PC and LSMO-AC. The minimum in the linewidth in panel (a) gives the temperature of the FM-PM phase transition.

### 3.4 Magnetocaloric effect

Magnetocaloric effect (MCE) is a  material property. When a magnetic field is applied to a sample, it causes a change in magnetic state as well as structural rearrangement, resulting in a change in magnetic entropy, $\Delta S_M$ [4]. An indirect method was used to study the MCE, which is based on the thermodynamic Maxwell relation between the change in entropy, $S$, of magnetic material when a magnetic field $H$ is applied and the temperature dependence of magnetization [4]:

$$\left(\frac{\partial S}{\partial H}\right)_T = -\left(\frac{\partial M}{\partial T}\right)_H \tag{4}$$

The magnetic entropy change $\Delta S_M$ and the magnetic adiabatic temperature change $\Delta T_M$ are calculated as:

$$\Delta S_M(T, H) = \int_{H_1}^{H_2} \left(\frac{\partial M}{\partial T}\right)_H dH \tag{5}$$

$$\Delta T_M(T,H) = \int_{H_1}^{H_2} \frac{T}{C_P}\left(\frac{\partial M}{\partial T}\right)_H dH \qquad (4)$$

Here $H_2$–$H_1$ is the change of the applied field and $c_p$ is the specific heat of the sample.

The variation of $\Delta S_M$ as a function of temperature at different applied magnetic fields for both samples is shown in **Fig.9 (a)** and **(b)**. The values for $\Delta S_M$ are negative across the entire temperature range, indicating that the adiabatic change in the magnetic field is accompanied by the release of heat, confirming the ferromagnetic behavior of our systems [4]. The magnetic entropy change reaches its maximum at $T_C$ and increases with applying the magnetic field. The maximum value of magnetic entropy change, $-\Delta S_M^{max}$ is found to be 1.69 J/kg.K around 340 K for LSMO-PC and 1.09 J/kg.K around K for LSMO-AC in a field of 5 T. Many elements, including surface effects, finite size effects, defect expansion, an increase in cationic vacancies, and the existence of broken bonds on particle surfaces, can explain the decrease in maximum entropy change [62]. These common effects in fine particle samples lead to a reduction in FM interaction. In this case, particles of smaller size contribute less to the magnetic entropy change [63]. It can also be seen that the magnitude of the $\Delta S_M$ peaks gradually broadens as the grain size decreases [64]. The magnitude, temperature dependency, and magnetic field dependence of the magnetic entropy change depend significantly on the nature of the magnetic phase transition. A second-order magnetic phase transition is observed in both samples LSMO-PC and LSMO-AC. The first-order transition takes place in a narrow coexistence range, thus limiting the MCE in a narrow temperature range. The second-order transition, on the other hand, occurs across a large temperature range, which is advantageous for active magnetic regenerative [4]. Because of the magnetic inhomogeneity in the samples, the ferromagnetic transition is broadened, and, as a result, the -$\Delta S_M$ decreases but spreads over a wide temperature range. This may explain the low -$\Delta S_M$ values in LSMO-AC. The grain size, and more precisely, the surface-to-volume ratio, can explain the broadening of the $\Delta S_M$ peak [64], [65]. The proportion of surface area increased when the crystallite size was reduced, resulting in a weaker FM coupling to the core [66]. As a result, the Curie temperature values became dispersed, resulting in a broad magnetic transition. Furthermore, the strong enhancement of the magnetocaloric effect in LSMO-PC is most likely due to the formation of ferromagnetic tendency at the surface of the nanoparticles present [67].

**Fig.10 (a)** and **(b)** display the variation of the magnetic adiabatic temperature change $\Delta T_M$ as a function of temperature at different applied magnetic fields for LSMO-PC and LSMO-AC, respectively. The $\Delta T_M$ reveals the same behaviors as $\Delta S_M$. Around $T_c$, $\Delta T_M$ exhibits a maximum, described by equation (4). As shown in **Fig.10**, the $\Delta T_M^{max}$values were 1.04 K, around 343K for LSMO-PC, and 0.59 K, around 313K for LSMO-AC. These values are comparable with that of the other manganites [4]. For $La_{0.7}Sr_{0.2}MnO_3$, the values of $\Delta T_M$=1.01K and3.33Kfor $\Delta H$= 1 T and 6 T at 365 K, respectively, were reported [18]. It is evident that materials with a more significant magnetization gradient and a higher saturation field will show larger $\Delta T_M$.

Around the Curie temperature, the dependency of the magnetic field of $\Delta S_M^{max}$ for materials with a second-order transition can be study. **Fig.11** displays the magnetic field dependence of $\Delta S_M^{max}$ for LSMO-PC and LSMO-AC. The fitting of the curves with the power law gives a critical exponent $n \approx 0.88$ and $n \approx 0.87$ for LSMO-PC and LSMO-AC, respectively. The values of $n$ are higher than what was expected by mean-field theory, where $n$ is typically equal to 2/3 around $T_C$ [68]. This difference confirm that both of LSMO-PC and LSMO-AC are not completely paramagnetic at this temperature and short-range ferromagnetic order in the paramagnetic phase appears.

To evaluate the applicability of LSMO-PC and LSMO-AC as magnetic refrigerants, the obtained values of $\Delta S_M^{max}$and $\Delta T_M^{max}$ are compared in **Table 2** with those reported in the literature for several other magnetic materials. D.N.H.Nam et al. reported the value of $-\Delta S_M^{max}$= 1.64 J/kg.K and $\Delta T_M^{max}$ = 1.01 K in $La_{0.7}Sr_{0.3}MnO_3$ for $\Delta H$ = 7 T at 364 K [69]. The MCE in perovskite manganites is generally caused by a large variation in magnetization near the magnetic phase transition [3]. The $La_{0.75}K_{0.2}Mn_{1.05}O_3$ sample shows a larger magnetic entropy change $-\Delta S_M^{max}$ = 3.629 J/kg.K under $\Delta H$ = 3 T at $T_C$ = 332 K [70]. The DE mechanism impacts the MCE. The process of changing magnetic entropy involves spin-lattice and spin-orbit coupling. Therefore, an essential factor in the variation in magnetization near $T_C$ is the change in Mn-O bond length, Mn-O-Mnangle, unit cell volume, and crystalline phase [34], [71], [72]. Arun et al. elaborated the $Nd_{0.50}\square_{0.17}Sr_{0.33}MnO_3$ monocrystalline by sol-gel reaction, revealing that the MCE response reached a value of $-\Delta S_M^{max}$= 0.98 J/kg.K under $\Delta H$= 5 T at 266 K, which is lower than our results [73]. The obtained values and the advantages of perovskite manganites, such as high yield, good chemical stability, high electrical resistivity, and low production costs, make LSMO-PC and LSMO-AC promising candidates for eco-friendly room temperature magnetocaloric applications.

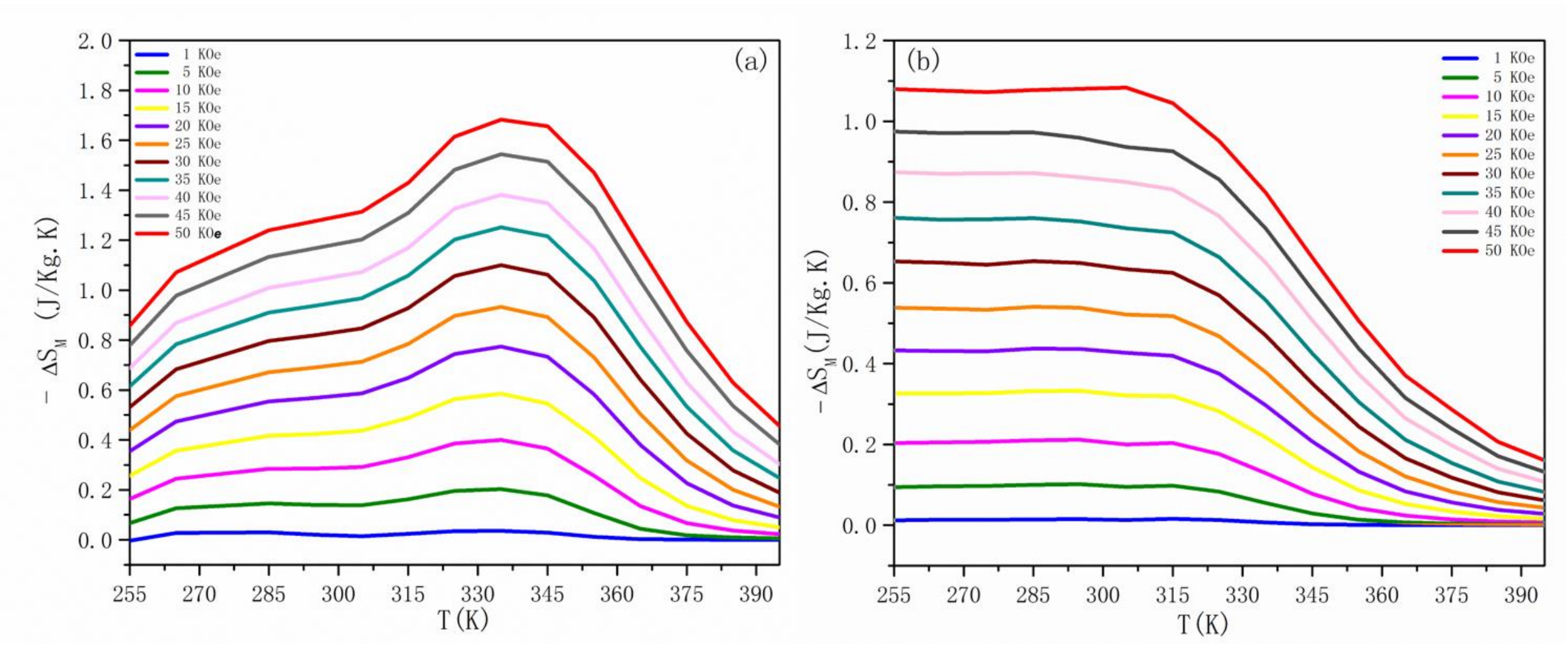

Figure 9: Magnetic entropy change, $\Delta S_M$, as a function of temperature at different applied magnetic fields in (a) LSMO-PC and (b) LSMO-AC.

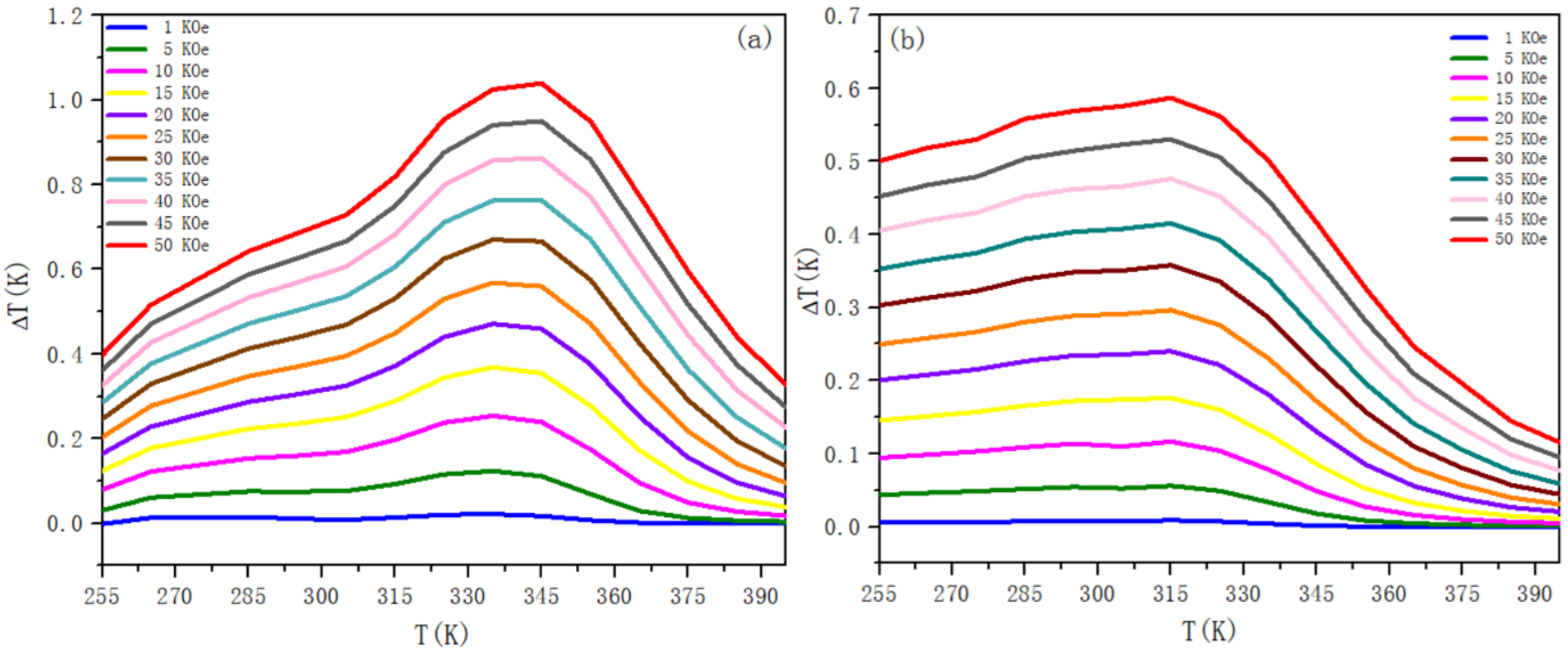

Figure 10: The magnetic adiabatic temperature change $\Delta T_M$ as a function of temperature at different applied magnetic fields for (a) LSMO-PC and (b) LSMO-AC.

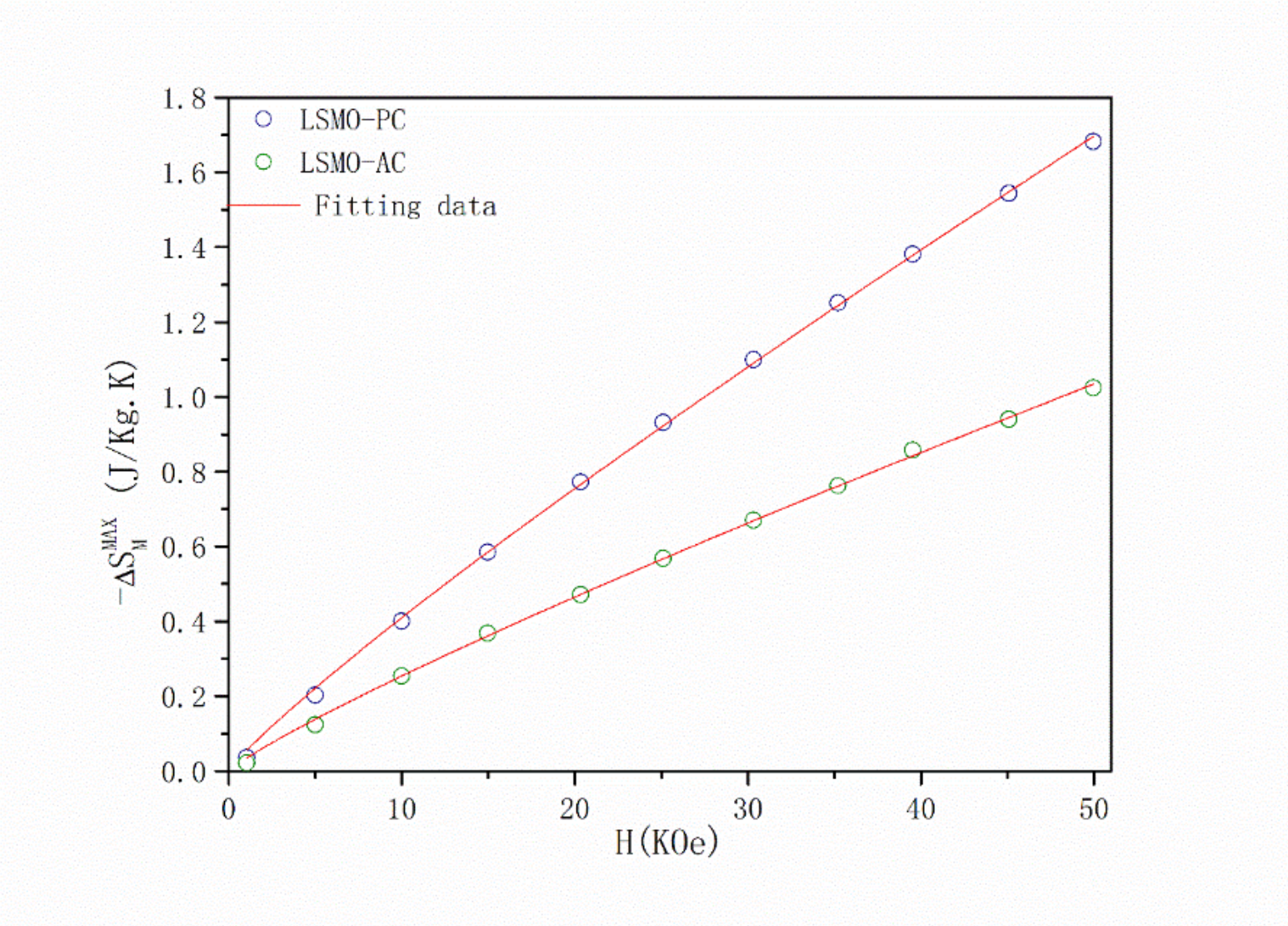

Figure 11: Maximum entropy change as a function of applied field for LSMO-AC and LSMO-PC.

Table 2: Magnetocaloric properties of ferromagnetic oxides.

| Materials | $T_C$ (K) | ΔH (T) | $\Delta T_M^{max}$ (K) | $-\Delta S_M^{max}$ (J/kg.K) | Refs. |
|---|---|---|---|---|---|
| $La_{0.8}Sr_{0.2}MnO_3$ (LSMO-PC) | 340 | 5 | 1.04 | 1.69 | This work |
| $La_{0.8}Sr_{0.2}MnO_3$ (LSMO-AC) | 290 | 5 | 0.59 | 1.09 | This work |
| $La_{0.7}Sr_{0.3}MnO_3$ | 365 | 1 | 1.01 | 1.64 | [69] |
| $La_{0.75}K_{0.2}Mn_{1.05}O_3$ | 332 | 3 | - | 3.629 | |
| $La_{0.67}Sr_{0.33}MnO_3$ | 370 | 5 | 3.33 | 5.15 | [75] |
| $La_{0.67}Sr_{0.33}MnO_3$ | 370 | 2.5 | 1.53 | 2.41 | [75] |
| $Nd_{0.50}\square_{0.17}Sr_{0.33}MnO_3$ | 266 | 5 | - | 0.98 | [73] |
| $La_{0.7}Ca_{0.2}Ba_{0.1}MnO_3$ | 230 | 2 | | 1.5 | [76] |

## 4 Conclusions

$La_{0.8}Sr_{0.2}MnO_3$ (LSMO) powders are synthesized using two distinct methods: Pechini (LSMO-PC) and auto-combustion (LSMO-AC). According to the XRD analysis, the material crystallized in mixed crystalline phases with rhombohedral and orthorhombic symmetries. The most prevalent crystallographic phase appears to be the rhombohedral phase, with ~ 92.76 % and ~ 98.10% for LSMO-PC and LSMO-AC, respectively. LSMO-PC is more likely to participate in the DE mechanism than LSMO-AC due to its higher proportion of orthorhombic phase (~ 7.24 %), leading to stronger ferromagnetic interactions and a higher curie temperature. According to SEM images, the average values of particle sizes are around 495 nm and 195 nm for LSMO-PC and LSMO-AC, respectively. Magnetic measurements confirm that the samples have ferromagnetic properties with Curie temperature $T_C$ =340 K and 290 K for LSMO-PC and LSMO-AC, respectively. LSMO-AC and LSMO-PC samples are close to conventional universality class presenting a small deviation from the mean-field values in opposite direction. The LSMO-PC has interesting magnetic and magnetocaloric properties. The LSMO-PC sample had the highest magnetization value of 60.82 emu/g. The change in magnetic entropy $-\Delta S_M$ measured indirectly via the Maxwell approach for the LSMO-PC increases with increasing field change ΔH and reaches 1.69 J/kg.K at 340 K for ΔH = 5 T. The corresponding adiabatic temperature change $\Delta T_M^{max} =$ 1.04 K is somewhat larger than $\Delta T_M^{max} = 0.59$ K observed in LSMO-AC. In addition, the porous surface of LSMO-PC allows faster heat exchange with regenerator fluid in cooling devices exploiting the active regeneration technique. On the other hand, LSMO-AC displays very high thermal stability of both $\Delta T_M$ and $\Delta S_M$. The results highlight the effect preparation method on the magnetocaloric performances of $La_{0.8}Sr_{0.2}MnO_3$ and show that both LSMO-AC and LSMO-PC compounds could be suitable candidates for room-temperature magnetic refrigeration applications.

**CRediT author statement**

**Afaak Lakouader:** Investigation, Methodology, Data Curation, Writing - Original Draft, Validation. **Youness Hadouch:** Visualization, Writing – review & editing, Validation. **Daoud Mezzane:** Visualization, Methodology, Writing – review & editing, Validation, Supervision. **Valentin Laguta:** InvestigationWriting – review & editing, Visualization, Validation. **M'barek Amjoud:** Visualization, Writing – review & editing, Validation. **Voicu O. Dolocan:** InvestigationWriting – review & editing, Visualization, Validation**. Nikola Novak:** Visualization, Writing – review & editing, Validation. **Lahoucine Hajji:** Visualization, Validation. **El Hassan Choukri:** Visualization, Writing – review & editing, Validation, Supervision. **Anna Razumnaya:** Visualization, Validation. **Abdelhadi Alimoussa:** Formal analysis**. Kutnjak** Visualization, Writing – review & editing, Validation**. Igor A. Luk'yanchuk:** Visualization, Writing – review & editing, Validation, Supervision. **Mimoun El Marssi:** Visualization, Methodology, Writing – review & editing, Validation, Supervision.

**Declaration of Competing Interest**

The authors declare that they have no known competing financial interests or personal relationships that could have appeared to influence the work reported in this paper.

**Data availability**

No data was used for the research described in the article

**Funding**

This work was supported by the European Union Horizon 2020 Research and Innovation actions MSCA-RISE-ENGIMA (No. 778072), MSCA-RISE-MELON (No. 872631).

**Acknowledgment**

This research is financially supported by the European Union Horizon 2020 Research and Innovation actions MSCA-RISE-ENGIMA (No. 778072), MSCA-RISE-MELON (No. 872631), and the Slovenian Research Agency grant P1-0125. We kindly thank Mathieu Rouzières (CRPP - CNRS Pessac) for helping with SQUID measurements.